\documentclass[12pt]{iopart}
\usepackage{epsf}
\usepackage{epsfig}
\usepackage{amssymb}

\begin{document}

\title[Aging of the KPZ equation]{Aging dynamics of non-linear elastic interfaces: the Kardar-Parisi-Zhang equation}

\author{Sebastian Bustingorry}
\address{DPMC-MaNEP, Universit{\'e}
de Gen{\`e}ve, 24 Quai Ernest Ansermet, 1211 Gen{\`e}ve 4, Switzerland}

\begin{abstract}
In this work, the out-of-equilibrium dynamics of the Kardar-Parisi-Zhang equation in (1+1) dimensions is studied by means of numerical simulations, focussing on the two-times evolution of an interface in the absence of any disordered environment. This work shows that even in this simple case, a rich aging behavior develops. A multiplicative aging scenario for the two-times roughness of the system is observed, characterized by the same growth exponent as in the stationary regime. The analysis permits the identification of the relevant growing correlation length, accounting for the important scaling variables in the system. The distribution function of the two-times roughness is also computed and described in terms of a generalized scaling relation. These results give good insight into the glassy dynamics of the important case of a non-linear elastic line in a disordered medium.
\end{abstract}


\maketitle


\section{Introduction}
\label{sec:intro}

Due to its relevance on fundamental problems of condensed matter physics, the statics and dynamics of elastic manifolds are a subject of great current interest. In
general, elastic manifolds can be related to problems such as domain wall motion in magnetic~\cite{magnetic} or ferroelectric~\cite{ferroelectric} materials,
vortex matter in high temperature superconductors~\cite{rusos,vortex,us}, grain boundary fluctuations in materials science~\cite{grains}, interface dynamics in deposition problems~\cite{Barabasi-Stanley}, or crack propagation~\cite{cracks}. The main
ingredients characterizing these problems might include the elastic energy of
the internal degrees of freedom, the quenched disorder, and the interaction between
different components of the system. It is known that when quenched disorder is present, these systems presents glassy characteristics such as disorder roughness exponents or non-stationary dynamics. In order to better understand such features as the slow out-of-equilibrium dynamics, it is important to study in the first place the properties of simpler models, i.e. systems without disorder and/or interactions.

It has been shown that even the simple Edwards-Wilkinson equation, a very well studied model of interface dynamics, presents glassy behavior~\cite{Cukupa,Pleimling,cuentas_EW}. Since this is an harmonic solvable model, it is possible to obtain from it a lot of information about the out-of-equilibrium dynamics, such as the multiplicative aging scaling of the two-times roughness,
the inclusion of finite size equilibration, and the generalization of scaling distribution functions. It is important to understand how these properties, obtained for the simple EW equation, can be generalized to other models as the non-linear Kardar-Parisi-Zhang (KPZ) equation~\cite{KPZ}, different discrete models~\cite{Barabasi-Stanley}, and even different dimensions. For these cases, however, the universal exponents characterizing static and dynamic properties might change, and then it becomes important to understand how such a change affects the glassy properties.

In particular, the study of the non-linear contributions to the elasticity should allow a better understanding of the glassy properties of more realistic elastic models. In this sense, the study of the out-of-equilibrium dynamics of the pure KPZ equation serve as a first step to rationalize the aging behavior reported in~\cite{Spaniards}, where a driven non-linear elastic model with quenched disorder was analyzed. Moreover, it shall also give a first insight into how the general picture obtained for the EW case can be generalized to more complex, and analytically non-solvable, models.

In addition, it is now clear how important fluctuations are to describe the out-of-equilibrium dynamics of complex systems~\cite{Bramwell,Racz,Chamon-etal,Chamon-Cugliandolo}. The thermally induced distribution functions of a given quantity contain more information than the first moments, which resulted in the proposal of using these fluctuations to characterize different universality classes~\cite{Racz,Aarao}. The roughness distribution of elastic lines was studied in the steady state for various models on the saturation regime~\cite{Aarao,Marinari,static-inter}, and also in the dynamic growth regime~\cite{Zoltan}. It was also analyzed in the context of depinning of elastic lines in random environments~\cite{Rosso} and in relation to the $1/f$ noise~\cite{gyorgyi}. Its generalization to non-equilibrium situations has also been recently considered~\cite{cuentas_EW,Bustingorryetal}. Within this context, it seems important to test the scaling of the out-of-equilibrium roughness distribution for the KPZ equation.

Therefore, with the aim of analyzing the glassy properties of non-linear elastic models, the numerical solution of the KPZ equation in (1+1) dimensions is presented in this work. Two-times correlation functions, as the roughness and structure factor, are computed and compared with the EW results. The distribution function of the roughness in the aging regime is also considered. These results allow to generalize those previously obtained for the EW equation by properly including the KPZ universal exponents. The paper is organized as follows. The numerical details of the simulations, together with the out-of-equilibrium protocol, are presented in \sref{sec:model}. The glassy properties for the different two-times quantities are presented and discussed in \sref{sec:results}. Finally, \sref{sec:conclusions} presents the conclusions.

\section{Numerical details}
\label{sec:model}

In the present section the numerical details of the simulation of the out-of-equilibrium dynamics of the KPZ equation are presented. This equation describes, in (1+1) dimensions, the evolution of a one-dimensional non-linear elastic object embedded in a two-dimensional space in the limit of small fluctuations. The non-linear contribution is a first order correction to elasticity, thus allowing the characterization of the line in terms of the univaluated field $x(z,t)$. The KPZ equation in (1+1) dimensions is then~\cite{KPZ}
\begin{equation}
\label{eq:kpz}
\frac{\partial x(z,t)}{\partial t} = \nu \frac{\partial^2 x(z,t)}{\partial z^2}+ \lambda \left[ \frac{\partial x(z,t)}{\partial z} \right]^2+\xi(z,t),
\end{equation}
where the thermal noise $\xi$ is characterized by
\begin{eqnarray}
\label{eq:noise}
\langle \xi(z,t) \rangle = 0,\\
\langle \xi(z,t) \xi(z',t') \rangle = 2\,T\, \delta(z-z')
\delta(t-t'),
\end{eqnarray}
with $\nu$ the elastic constant, $\lambda$ characterizing the strength of the non-linear term, $T$
the temperature of the thermal bath and $\langle \cdots \rangle$ the average
over the white noise $\xi$, {\it i.e.} the thermal average.

The KPZ equation~\eref{eq:kpz} is numerically solved using the following finite
difference approach:
\begin{equation}
\label{eq:disc-kpz}
\fl
x_i(t+t_0)=x_i(t)+t_0 \nu \left[ x_{i+1}(t)-2 x_i(t)+x_{i-1}(t) \right]+
\frac{t_0\,\lambda}{3} \Psi_i^2
+\sqrt{24 T t_0} \; \xi_i(t).
\end{equation}
Here, the internal dimension is replaced by a one-dimensional lattice of unitary space,
which sets the length scales in the following. The noise $\xi_i(t)$ is now a
random number uniformly distributed in $[-1/2,1/2)$. The particular choice of
the finite difference representation of the non-linear term is given by
\begin{equation}
\label{eq:disc-KPZ-term}
\fl
\Psi_i^2=\left[ x_{i+1}(t)-x_i(t)\right]^2 +\left[x_{i+1}(t)-x_i(t)\right]\left[x_i(t)-x_{i-1}(t)\right]+\left[x_i(t)-x_{i-1}(t)\right]^2,
\end{equation}
which not only gives the correct exponents in the KPZ universality
class but also the correct prefactors and the correct fluctuations in the saturation regime~\cite{Lam}. This is important in order to perform a direct comparison with the EW case. In this work, the parameters used to numerically solve the KPZ equation are $L=1024$ (except for the computation of distribution functions where the values $L=256$ and $L=64$ are also used in order to highlight the scaling properties), $t_0=0.01$, which sets the time units, $\nu=1$, and $\lambda=1$. For the thermal average, $N_T=10^3$ noise realizations were considered for the evaluation of two-times correlations. In order to compute the roughness distributions, $N_T=10^5$ thermal noise realizations were used for $L=256$ and $L=64$, while $N_T=10^4$ were used for $L=1024$.

The out-of-equilibrium dynamics of the KPZ equation was analyzed adopting the usual two-times protocol used to study glassy systems~\cite{Leto}. Starting from a flat configuration, the system is first equilibrated at a given initial temperature $T_0$. This means that the system evolves until the stationary state is reached, corresponding to the size-dependent saturation regime of the roughness. Then, the temperature is suddenly changed to the working temperature $T$, and the time is set to zero. Two-times correlation functions are then defined in terms of the waiting time $t_w$ and the elapsed time $\Delta t=t-t_w$. In the present work, the working temperature was set to $T=1$ and different initial temperatures, both smaller and larger than $T$, were studied. In particular, the values $T_0=0$, which corresponds to a perfectly flat initial condition, and $T_0=5$ were used.

\section{Results}
\label{sec:results}

In the following, the results for different two-times correlation quantities are described. Particular attention is given to correlation functions showing a clearly non-trivial out-of-equilibrium generalization of the scaling properties previously found in the EW equation.

\subsection{Roughness}
\label{sec:roughness}

The one-time roughness, which characterizes the width of the discrete fluctuating line, is commonly defined as~\cite{Barabasi-Stanley}
\begin{equation}
\label{eq:w2det}
w^2(t)=\frac{1}{L} \sum_{i=1}^L \left\langle \left[
\delta x_i(t) \right]^2 \right\rangle,
\end{equation}
where $\delta x_i(t)=x_i(t)-\overline{x(t)}$ and $\overline{x(t)}=L^{-1} \sum_{i=1}^L x_i(t)$ is the instantaneous center of
mass position. In general, the time evolution of the roughness can be described through
the Family-Vicsek scaling~\cite{Favi}, for which the roughness can be written as
$w^2(t) \sim L^\zeta f(t/t_x)$, with $t_x \sim L^z$ the saturation time and the scaling function $f(u)\sim u^\beta$ for $u \ll 1$ and $f(u) \sim const.$ for $u \gg 1$. This defines the
growth exponent $\beta$, the roughness exponent $\zeta$ and the dynamic
exponent $z$, related through the scaling relation $z=\zeta/\beta$. The values of these exponents are $\zeta=1$, $z=2$, and $\beta=1/2$ for the EW universality class, and $\zeta=1$, $z=3/2$, and $\beta=2/3$ for KPZ universality class, both in (1+1) dimensions.

\begin{figure}[!tbp]
\centerline{
\includegraphics[angle=-0,width=6cm,clip=true]{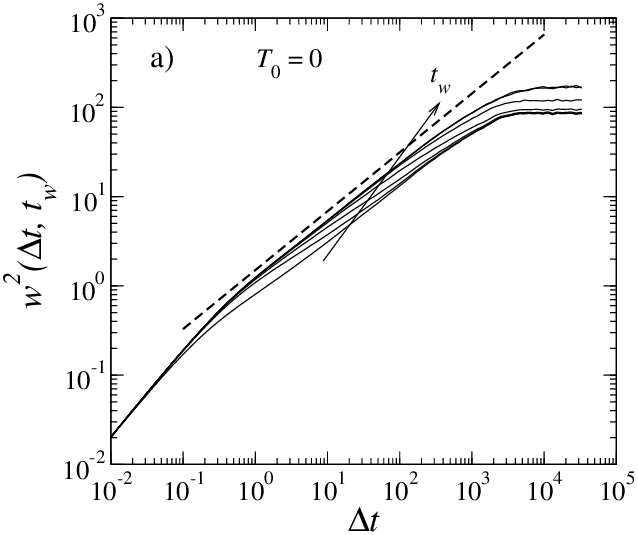}
\hspace{0.25in}
\includegraphics[angle=-0,width=6cm,clip=true]{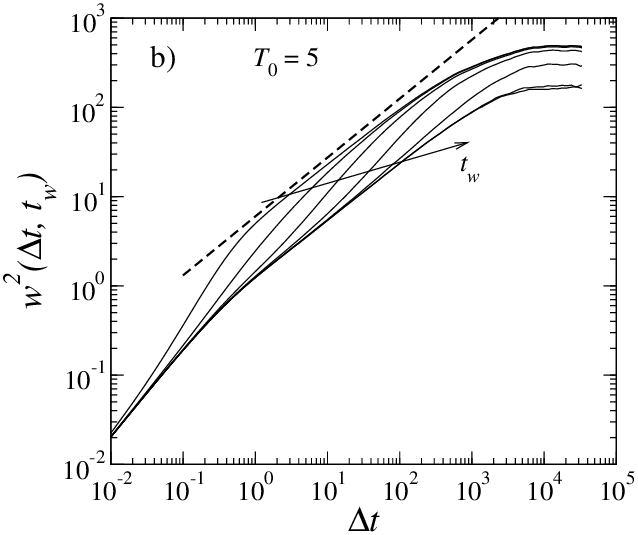}
} \caption{\label{f:w2_KPZ_1024}
Two-times evolution of the roughness for (a) $T_0=0$ and (b) $T_0=5$. The different waiting times are $t_w=0,\,1,\,8,\,64,\,512,\,4096$, and $32768$, with the arrow indicating how $t_w$ grows. For the two largest $t_w$ values the curves are superimposed, indicating that the result is stationary. The dashed line indicates the $\beta =2/3$ slope.}
\end{figure}

The two-times roughness, which is the out-of-equilibrium generalization of the one-time
roughness~\eref{eq:w2det}, is defined as
\begin{equation}
\label{eq:w2} w^2(\Delta t,t_w)=\frac{1}{L}\sum_{i=1}^L \left\langle \left[
\delta x_i(t_w + \Delta t) - \delta x_i(t_w) \right]^2 \right\rangle.
\end{equation}
This quantity measures the relative fluctuation between the line's configuration at times $t_w$ and $t$. It was recently shown that, for the EW equation with finite size $L$, the two-times roughness ages~\cite{cuentas_EW}, now scaling as $w^2 \sim F(t_w/t_x,\Delta t/t_x)$. The aging regime may be understood as the pre-asymptotic non-equilibrium regime prior to finite size equilibration, which takes a time $t_x \sim L^z$. When $t_w \gg t_x$, the stationary solution and the Family-Vicsek scaling are recovered. On the other hand, in the aging regime $t_w \ll t_x$, the scaling function behaves as
\begin{eqnarray}
\label{eq:gen-fv}
F\left(\frac{t_w}{t_x},\frac{\Delta t}{t_x}\right) \sim \left\{
\begin{array}{lll}
t_w^\beta f_1\left(\frac{\Delta t}{t_w}\right) & \mbox{for} & \Delta t \ll t_x,
\\
\\
L^\zeta f_2\left(\frac{t_w}{t_x}\right)& \mbox{for} & \Delta t \gg t_x,
\end{array}
\right.
\end{eqnarray}
with $\beta=1/2$ and $\zeta=1$ the EW scaling exponents. In the case where $\Delta t \ll t_x$, the roughness scales as $w^2 \sim t_w^\beta f_1(\Delta t/t_w)$, with $f_1(u) \sim c(T) u^\beta$ for $\Delta t \ll t_w$ and $f_1(u) \sim c_0(T_0,T) u^\beta$ for $\Delta t \gg t_w$. This scaling describes a multiplicative aging scenario~\footnote{Note that for $\Delta t \ll t_x$ the roughness can be also written as $w^2 \sim \Delta t^\beta f_1'(\Delta t/t_w)$, with $f_1' \sim c(T)$ for $\Delta t \ll t_x$ and $f_1'(u) \sim c_0(T_0,T)$ for $\Delta t \gg t_w$. \Eref{eq:gen-fv} emphasizes the fact that this corresponds indeed to multiplicative aging.} for the two-times roughness~\cite{Cukupa,cuentas_EW,Pleimling}, and effectively corresponds to the infinite size limit, $L \to \infty$. On the other hand, when $\Delta t \gg t_x$, i.e. in the saturation regime, the scaling function $f_2(t_w/t_x)$ behaves as $f_2(u) \sim s_0(T,T_0)$ for $t_w \ll t_x$ and $f_2(u) \sim s(T)$ for $t_w \gg t_x$, thus leading to a $t_w$-dependent saturation value for the roughness which evolves from $s_0(T,T_0)L^\zeta$ to $s(T)L^\zeta$ while $t_w$ increases. The scaling form \eref{eq:gen-fv} generalizes the Family-Vicsek scaling to a non-equilibrium situation. It is interesting to test if this generalization is valid for the KPZ equation, using its respective exponents. This point is addressed in the following.

\begin{figure}[!tbp]
\centerline{
\includegraphics[angle=-0,width=6cm,clip=true]{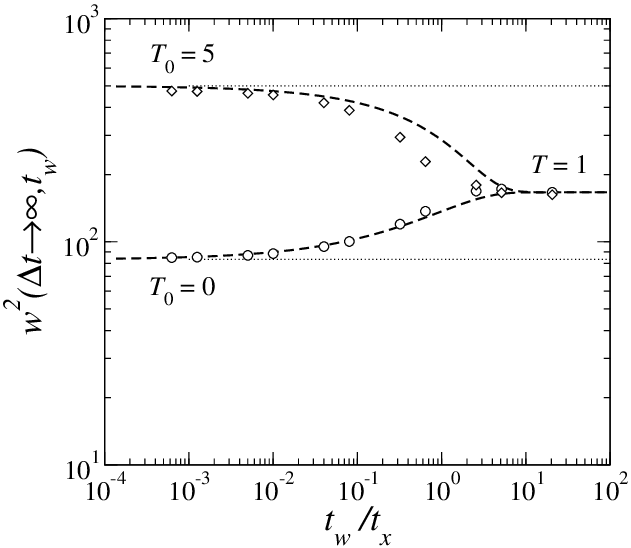}
} \caption{\label{f:w2_sattw}
Evolution of the saturation roughness with the waiting time for $T_0=0$ (circles) and $T_0=5$ (diamonds), corresponding to the numerical solution of the KPZ equation. The continuous dashed lines correspond to the EW equation analytical solution. Dotted lines indicate the $t_w \ll t_x$ limit $L^\zeta(T_0+T)/12$. For comparison, the waiting time is rescaled in terms of the saturation times, $t_x = L^2/144$ and $t_x = L^{3/2}/(15/2)^{3/2}$, for the EW and KPZ cases respectively.}
\end{figure}

In this work, the two-times roughness $w^2(\Delta t,t_w)$, has been numerically obtained from the evolution of the discrete KPZ equation, and it is shown in~\fref{f:w2_KPZ_1024}. Panels (a) and (b) correspond to initial conditions with $T_0=0<T$ and $T_0=5>T$, respectively. The different curves correspond to different waiting times, as indicated. After an initial time interval where the components of the line fluctuate independently, $w^2 \sim \Delta t$, the line becomes correlated in the longitudinal direction and starts aging. Then, ones the line is correlated, two different time regimes are observed, corresponding to $\Delta t$ larger or smaller than $t_x$, and both showing aging. When the waiting time is larger than the saturation time, aging stops and the stationary situation is reached.

In the growth regime, for each waiting time, the curves jump from the equilibrated asymptote to the non-equilibrated asymptote with increasing $\Delta t$. Both asymptotes are growing as $\Delta t^\beta$, with $\beta =2/3$ the KPZ value. Therefore, a multiplicative aging scenario with the corresponding KPZ growth exponent is found. In the saturation regime the curves jump from the $s_0(T,T_0)L^\zeta$ asymptotic value, corresponding to $t_w=0$, to the value $s(T)L^\zeta$, corresponding to the stationary solution. The values of the numerically obtained prefactors $s_0$ and $s$ are equal, within numerical accuracy, to those analytically found for the EW equation, i.e. $s_0=(T_0+T)/12$ and $s=T/6$. It is well known that the KPZ and EW continuum equations have the same steady state solution for the distribution function of a given configuration in the saturation regime~\cite{Barabasi-Stanley}. Furthermore, it has been also shown that this property holds for the discretized model used here~\cite{Lam}. However, the fact that the $t_w$-dependent saturation values obtained in this work coincide with the ones for the EW equation suggests that the non-stationary solution in the saturation regime may also be the same for both models. \Fref{f:w2_sattw} shows the $t_w$-dependent saturation value for both initial conditions, and compares it with the EW equation solution~\cite{cuentas_EW}. The asymptotic limits are clearly the same, and a small difference is observed in the crossover region.

These results support the generalized Family-Vicsek scaling, although the growth regime is too short in time to accurately test the $\Delta t/t_w$ scaling. Larger system sizes would be necessary for a good test using the two-times roughness. However, a better way to observe the growing correlation length associated with aging is to focus on the two-times structure factor. This quantity also reflects the scaling properties of the roughness, but a better scaling test is obtained, as described in the next section.

\subsection{Structure factor}
\label{sec:structure-factor}

\begin{figure}[!tbp]
\centerline{
\includegraphics[angle=-0,width=6cm,clip=true]{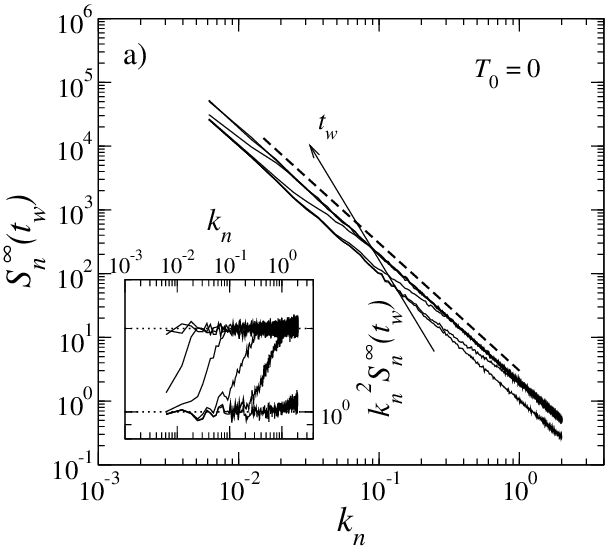}
\hspace{0.25in}
\includegraphics[angle=-0,width=6cm,clip=true]{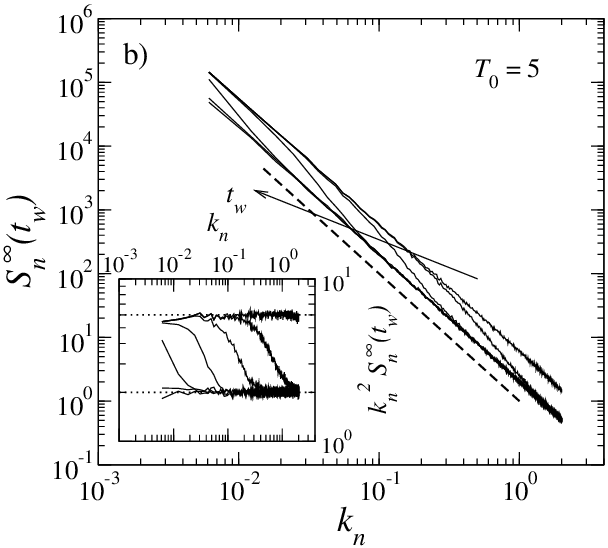}
} \caption{\label{f:Sn_KPZ_1024} Waiting time-dependent structure factor in the saturation regime, $S^\infty_n(t_w)$, for the initial conditions (a) $T_0=0$ and (b) $T_0=5$. The different waiting times are $t_w = 0,\,1,\,8,\,64,\,512,\,4096$, and $32768$. The dashed lines indicates the $k_n^{-2}$ behavior. The insets show a rescaled version of the main panels, using $k_n^2 S^\infty_n(t_w)$ against $k_n$ in order to emphasize the crossover between the two asymptotic values. Dotted lines in the insets correspond to the asymptotic value $T_0+T$ and $2T$. In the insets, $t_w$ increases from right to left.}
\end{figure}

The one-time dynamical structure factor is usually defined as
\begin{equation}
\label{eq:Sonet}
S_n(t)=L\langle \left| c_n(t) \right|^2 \rangle,
\end{equation}
where $c_n(t)$ represents the Fourier modes of a given configuration at time $t$, and where the discreetness of the lattice was already taken into account with the wave vectors given by $k_n=2 \pi n/L$, $n=1,...,L$. This structure factor gives information on how the fluctuations of the modes evolve in time, containing also information on the growing correlation length~\cite{Barabasi-Stanley,Kolton}. The definition \eref{eq:Sonet} should not be confused with the dynamical structure factor $S^D_n(t)=L\langle c_n(t) c_{-n}(0) \rangle$, which gives information on how a given mode correlates between times $t=0$ and $t>0$. This dynamical structure factor presents non-trivial stretched exponential relaxation characteristics in the
KPZ case~\cite{Eytan}.

In general, the structure factor \eref{eq:Sonet} scales as $S_n \sim t^{(1+\zeta)/z} g(k_n/k_x)$, with $k_x\sim t^{-1/z}$. The scaling function $g(u)$ behaves as $g(u) \sim const.$ for $k_n \ll k_x$ and $g(u) \sim u^{-(1+\zeta)}$ for $k_n \gg k_x$, indicating that there exists a growing correlation length $\ell \sim k_x^{-1} \sim t^{1/z}$. When $\ell \sim L$, the saturation regime is reached and $S_n^\infty \sim k_n^{-(1+\zeta)}$ is independent of the time $t$.

The two-times generalization of the structure factor is defined as~\cite{cuentas_EW}
\begin{equation}
\label{eq:SqDttw}
S_n(\Delta t,t_w)=L\langle \left| c_n(t_w+\Delta t)-c_n(t_w) \right|^2 \rangle.
\end{equation}
This definition allows for the characterization of the fluctuations of a given mode between $t_w$ and $t$. For the EW equation, the two-times structure factor can be written as a scaling function expressed in terms of the quotients $\Delta t/t_x$ and $t_w/t_x$, in analogy with the roughness scaling \eref{eq:gen-fv}. However, the focus here will be set on how the structure factor behaves on the $t_w$-dependent saturation regime, which gives clear evidence of the growing correlation lengths. Then, for $\Delta t \gg t_x$, i.e. in the saturation regime, the structure factor is computed for different $t_w$ values as
\begin{equation}
\label{eq:Sqtw}
S^\infty_n(t_w)=\lim_{\Delta t \gg t_x}L\langle \left| c_n(t_w+\Delta t)-c_n(t_w) \right|^2 \rangle.
\end{equation}
It was recently reported for the case of the EW equation that $S^\infty_n(t_w) \sim k_w^{-(1+\zeta)} g_1(k_n/k_w)$, with $k_w \sim t_w^{-1/z}$~\cite{cuentas_EW}. The scaling function $g_1(u)$ behaves as $g_1(u) \sim (T_0+T) u^{-(1+\zeta)}$ for $k_n \ll k_w$ and $g_1(u) \sim 2T u^{-(1+\zeta)}$ for $k_n \gg k_w$. The particular wave vector $k_w$ precisely separates two regimes. While the regime with large wave vectors, $k_n \gg k_w$, is equilibrated at the working temperature, small wave vectors, $k_n \ll k_w$, still have memory of the initial temperature. Note that in contrast with the one-time structure factor \eref{eq:Sonet}, the two-times generalization contains information on the dynamic exponent in the saturation regime due to its $t_w$-dependence.

\begin{figure}[!tbp]
\centerline{
\includegraphics[angle=-0,width=6cm,clip=true]{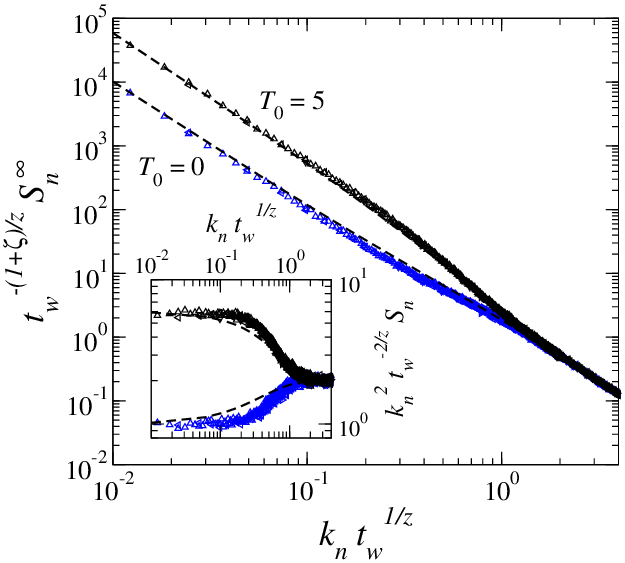}
} \caption{\label{f:Sn_KPZ_1024_scal} Scaling of the structure factor presented in~\fref{f:Sn_KPZ_1024} with $t_w^{-(1+\zeta)/z} S^\infty_n(t_w) \sim g(k_n\;t_w^{1/z})$. The solution of the EW equation is also shown as dashed lines. The inset shows the same data in the scaled form $k_n^2 t_w^{-2/z} S^\infty_n(t_w) \sim g(k_n\;t_w^{1/z})$.}
\end{figure}

\Fref{f:Sn_KPZ_1024} shows the structure factor $S^\infty_n(t_w)$ as a function of $k_n$ for different waiting times, obtained in the present work. Panels (a) and (b) correspond to $T_0=0$ and $T_0=5$, respectively. The structure factor clearly presents the $k_n^{-(1+\zeta)}$ decay, but changes between two asymptotes at a given $t_w$-dependent value. In order to emphasize this behavior, the insets show the same data plotted as $k_n^2\,S^\infty_n(t_w)$ against $k_n$. The crossover between the two asymptotes is the kind of behavior described by the scaling function $g_1(u)$. Indeed, the two asymptotic values are $T_0+T$ and $2T$ respectively, indicated with dotted lines in the insets. These values correspond to those previously obtained for the EW equation, again suggesting a strong connection between the non-stationary solution of the EW and KPZ equations in the saturation regime. This feature is stressed in \fref{f:Sn_KPZ_1024_scal}, where the scaling of the structure factor with $k_n/k_w \sim k_n\;t_w^{1/z}$ is shown for the two initial conditions. The data are also compared with the EW equation solution~\cite{cuentas_EW}. It can be seen that although the asymptotic values coincide, the intermediate dynamic regime does not perfectly match. This might be related to the fact that the dynamic exponent is also involved in the scaling function $g(u)$, which goes beyond both models having the same roughness exponent in (1+1) dimensions.

\subsection{Scattering function}
\label{sec:scattering-function}

Another quantity which has proved to be very useful in the analysis of the non-equilibrium dynamics in systems of interacting particles is the incoherent scattering function~\cite{Barrat,Bonn,Tanaka}. In the context of elastic lines it can be defined as~\cite{cuentas_EW}
\begin{equation}
\label{eq:scater}
C_q(\Delta t,t_w)=\frac{1}{L}
\sum_{i=1}^L \left\langle e^{iq\left[ \delta x_i(t_w + \Delta t)-\delta x_i(t_w) \right]} \right\rangle.
\end{equation}
It has been shown that due to the gaussian character of the variables involved in the solution of the EW equation, the scattering function can be written in terms of the two-times roughness in a very simple way as
\begin{equation}
\label{eq:scater-gauss}
C_q(\Delta t,t_w)=e^{-\frac{q^2}{2}\,w^2(\Delta t,t_w)},
\end{equation}
which establishes a clear relation between scattering functions and diffusion-like correlations. In the growth regime, $\Delta t \ll t_x$, it has been already shown that the roughness behaves as $w^2 \sim t_w^\beta f_1(\Delta t/t_w)$, with $f_1(u) \sim c(T) u^\beta$ for $\Delta t \ll t_w$ and $f_1(u) \sim c_0(T_0,T) u^\beta$ for $\Delta t \gg t_w$, leading to a stretched exponential relaxation of the scattering function, i.e.
\begin{equation}
\label{eq:scater-stretch}
C_q(\Delta t,t_w)=
\left\{
\begin{array}{lll}
e^{-\frac{1}{2}c(T)\,q^2\Delta t^\beta} & \mbox{for} & \Delta t \ll t_w,
\\
e^{-\frac{1}{2}c_0(T_0,T)\,q^2\Delta t^\beta} & \mbox{for} & \Delta t \gg t_w.
\end{array}
\right.
\end{equation}
This equation involves two stretched exponentials sharing the same exponent $\beta$ but different prefactors. This relation trivially holds for gaussian variables, however its validity for the KPZ equation must be still tested.

\begin{figure}[!tbp]
\centerline{
\includegraphics[angle=-0,width=6cm,clip=true]{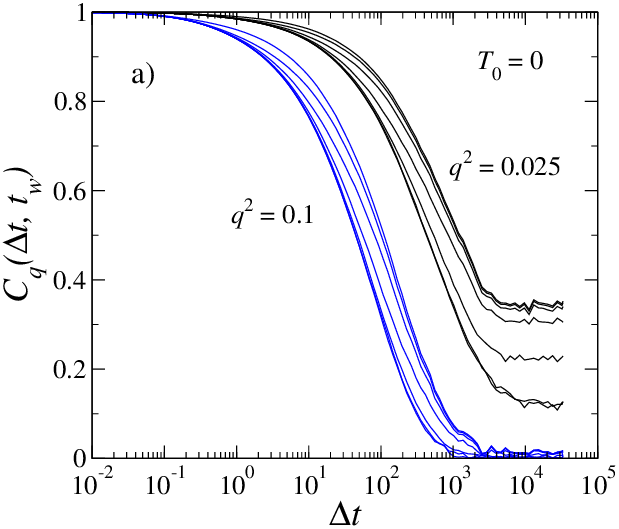}
\hspace{0.25in}
\includegraphics[angle=-0,width=6cm,clip=true]{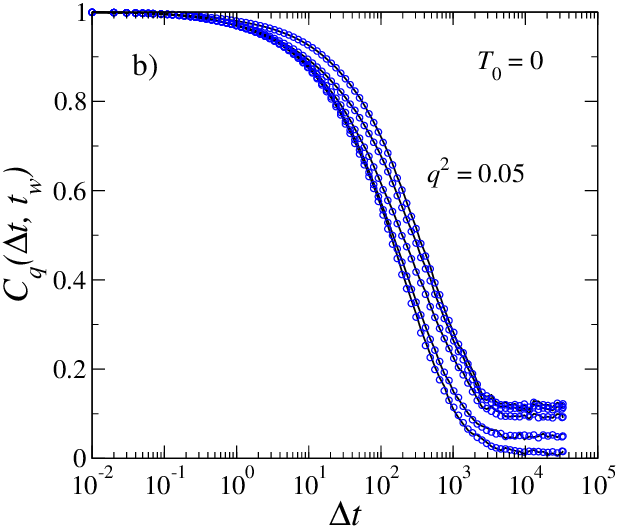}
} \caption{\label{f:Cq_KPZ_1024} (a) Two-times evolution of the scattering function for $T_0=0$. The different waiting times are $t_w =0,\,1,\,8,\,64,\,512,\,4096$, and $32768$ from upper to lower curves. For the two largest $t_w$ values the curves are superimposed, indicating that the result is stationary. The values of $q^2$ are indicated for each set of curves. (b) Comparison of the two-times evolution of the scattering function (black lines) with the value obtained from the roughness and using equation~\eref{eq:scater-gauss} (circles). The waiting times used here are $t_w = 0,\,8,\,64,\,512$, and $32768$, from upper to lower curves.}
\end{figure}

\Fref{f:Cq_KPZ_1024}(a) presents the scattering function for the KPZ equation obtained numerically at different $q$ and $t_w$ values. The scattering function presents a strong $q$-dependent saturation regime, on which the $q^2$ factor of the exponential competes with the $L^\zeta$ factor coming from the saturation regime of the roughness. Therefore, at low $q$ values a saturation regime can be observed, while at large $q$ values the scattering function decays faster to a near-zero value. This is clearly observed for the selected values $q^2=0.025$ and $q^2=0.1$ in \fref{f:Cq_KPZ_1024}(a). The generic form of the $C_q$ curves is similar to the one obtained for the EW equation. In order to test the validity of \eref{eq:scater-gauss}, the scattering function for $q=0.05$ is plotted in \fref{f:Cq_KPZ_1024}(b) together with the values $e^{-\frac{q^2}{2}\,w^2}$ obtained using the results for the roughness presented in~\fref{f:w2_KPZ_1024}(a). The results in \fref{f:Cq_KPZ_1024}(b) prove that relation~\eref{eq:scater-gauss} also holds for the KPZ equation. One could have expected this relation to hold in the saturation regime, since the steady state solution is the same for EW and KPZ equations. However, the fact that it also holds in {\it all} the dynamic $\Delta t$ and $t_w$-dependent growth regime is absolutely non-trivial, since the fluctuations of the KPZ are not expected to be gaussian distributed. The same results, i.e. the scattering function being directly obtained from the roughness through~\eref{eq:scater-gauss}, were obtained for $T_0=5$ (not shown here).

\subsection{Roughness distribution}
\label{sec:roughness-distribution}

It has been previously proposed that the roughness distribution function in the saturation regime scales with the average roughness as the only scaling parameter~\cite{static-inter}. This proposition can be extended to all the dynamic regime in steady state, allowing to write the roughness distribution as~\cite{Zoltan}
\begin{equation}
\label{eq:Phi-AntalRacz} P\left( w^2;t \right)=\frac{1}{w^2(t)}\;\Phi \left[
\frac{w^2}{w^2(t)}; \frac{t}{L^z} \right].
\end{equation}
Here, the fluctuating roughness $w^2$ should be distinguished from its average value $w^2(t)$, which explicitly contains the time dependence. The last argument in \eref{eq:Phi-AntalRacz} states that scaling works for curves with the same time scale $t/t_x$, thus accounting for finite size effects. This scaling relation can also be generalized to include the waiting time dependence by considering also the $t_w/t_x$ scale~\cite{us,cuentas_EW}, thus leading to
\begin{equation}
\label{eq:Phi-Dttw} P\left( w^2;\Delta t,t_w\right)=\frac{1}{w^2(\Delta
t,t_w)}\;\Phi\left[ \frac{w^2}{w^2(\Delta t,t_w)}; \frac{\Delta t}{L^z},
\frac{t_w}{L^z}\right].
\end{equation}
This last scaling relation was analitically obtained for the EW equation~\cite{cuentas_EW} and numerically obtained for a directed polymer in random media model~\cite{Bustingorryetal}.

\begin{figure}[!tbp]
\centerline{
\includegraphics[angle=-0,width=6cm,clip=true]{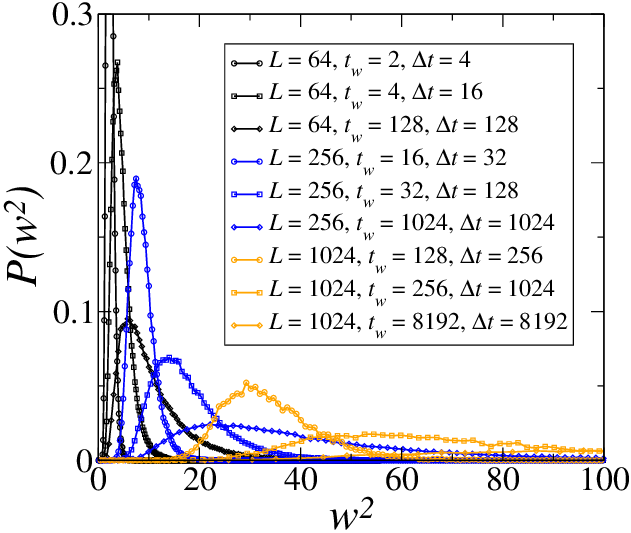}
\hspace{0.25in}
\includegraphics[angle=-0,width=6cm,clip=true]{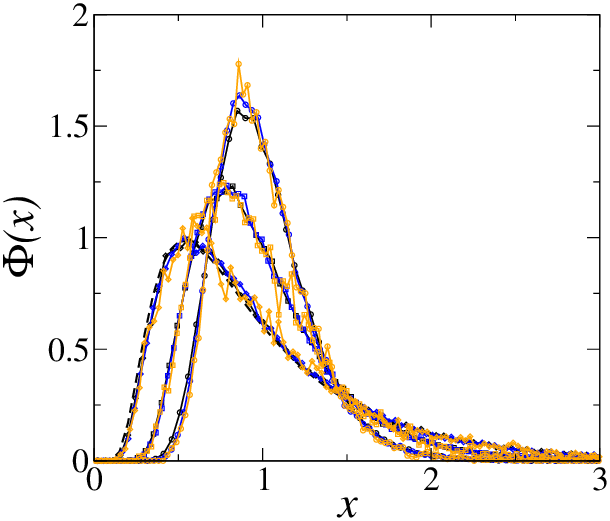}
} \caption{\label{f:Fw2_KPZ_scal} (a) The bare distribution function $P(w^2)$ for different system sizes $L$, waiting time $t_w$, and elapsed time $\Delta t$, as indicated. The initial temperature is $T_0=0$. (b) Scaled roughness distributions $\Phi(x)$, with $x=w^2/w^2(\Delta t,t_w)$. Symbols as in panel (a). Different sets of curves correspond to the same pair of values $(t_w/L^z,\Delta t/L^z)$. The dashed lines correspond to the EW solution in the stationary regime~\cite{static-inter}.}
\end{figure}

Using the results obtained in the present work a test of the scaling relation~\eref{eq:Phi-Dttw} for the KPZ equation can be performed. The result is shown in~\fref{f:Fw2_KPZ_scal}, obtained using the roughness distribution for three system sizes, $L=64$, $L=256$ and $L=1024$, and different times $\Delta t$ and $t_w$. In panel (a), the bare distribution functions are shown, while panel (b) shows the scaling function $\Phi(x)$, with $x=w^2/w^2(\Delta t,t_w)$. These results correspond to $T_0=0$. Since the difference between the selected system sizes is a factor four, the saturation time is scaled by a factor $4^z=8$. Then to keep the quotient $t_w/t_x$ fixed, the waiting time should also be scaled by the same factor $4^z=8$; the same holds for the value of $\Delta t$ if one wants to scale with $\Delta t/t_x$. It is shown that all the curves in panel (a) collapse into three sets of curves in panel (b), corresponding to different pairs of $(t_w/t_x,\Delta t/t_x)$ values, in agreement with the scaling relation~\eref{eq:Phi-Dttw}. This scaling of the roughness distribution function also accounts for the saturation regime, leading to the known stationary solution. A direct comparison with the EW results in the dynamic regime is not possible here because the scaling factor is different; in the present case it should be $4^z=16$ with $z=2$ for EW. Instead, the stationary saturation solution of the EW case~\cite{static-inter} is plotted in panel (b), with a continuous dashed line, showing again that the numerical solution of the KPZ equation asymptotically tends to the same distribution function in (1+1) dimensions\cite{Racz}.

\section{Conclusions}
\label{sec:conclusions}

The out-of-equilibrium dynamics of the KPZ equation has been studied in detail using numerical simulations. It corresponds to the relaxation of a non-linear interface from a given initial condition and it has been shown here that a complicated glassy dynamics emerges, as also does for the EW equation~\cite{cuentas_EW}. The system size $L$ was incorporated in the analysis, thus allowing to reach finite size equilibration after a time $t_x$. Therefore, the out-of-equilibrium regime described here effectively corresponds to a pre-asymptotic regime before equilibrium is reached. One can imagine a very large system size whose equilibration time $t_x$ is much larger than the observation time, making the aging regime the relevant observable time regime. In a sense this is expected when quenched disorder is taken into account, where the dynamics becomes much slower and the equilibration time goes beyond the observation time. Thus, the main aging characteristics described here are of key importance for analyzing the non-equilibrium dynamics of problems which include other components like disorder, external forces or line-line interactions.

It has been shown here that the main characteristics of the out-of-equilibrium dynamics of the EW equation can be extended to the KPZ equation. One should only be careful of using the proper scaling exponents. For instance, from the scaling of the two-times roughness and structure factor the existence of the growing correlation length is revealed, which allows to rationalize the different time regimes. In terms of elapsed time $\Delta t$ and waiting time $t_w$, it is the relative value of the correlation lengths $\ell (\Delta t)$ and $\ell(t_w)$ which defines the scaling properties in the growth regime. When considering the saturation regime, the size of the system $\ell(t_x) \sim L$, associated to the saturation time $t_x$, should also be considered. Thus, it is the competition between these three length scales, $\ell (\Delta t)$, $\ell(t_w)$ and $\ell(t_x)$, and its relation to the corresponding times through the dynamic exponent $z$, which define the aging dynamics of the system.

Results concerning the scattering function have been also presented. This correlation function presents a well defined stretched exponential decay, which is indeed commonly observed in different glassy systems~\cite{Leto}. Moreover, it was shown that the scattering function is directly related to the roughness through equation~\eref{eq:scater-gauss}. This simple exponential relation was previously analytically obtained for the EW equation based on the gaussian character of the fluctuations~\cite{cuentas_EW}. Although in the saturation regime the fluctuations in the KPZ equation are also gaussian distributed, in the growth regime this is not necessarily true. Thus, the fact that both scattering function and roughness are related in this simple way for the whole two-times dependent dynamics is a non-trivial result, posing new questions about the relation between scattering functions and displacements fluctuations out of equilibrium, a fact already pointed out in colloidal glass experiments~\cite{Bonn}.

Finally, the scaling of the roughness distribution has been studied. The scaling in equation~\eref{eq:Phi-Dttw} for the distribution function, which depends on the relative time scales $\Delta t/L^z$ and $t_w/L^z$, has been tested here and a good collapse of the data was obtained. This indicates once more that the correct variables which include both dynamics and finite size effects are the relative scales between $\ell (\Delta t)$, $\ell(t_w)$ and $\ell(t_x)$.

Therefore, it has been highlighted here that the scaling relations analytically found for the EW equation are quite robust, allowing also for a good description of the out-of-equilibrium dynamics of the KPZ equation. While the present work focused on the numerical solution of the continuum equation, it could also be interesting to test these ideas in discrete models of interface dynamics. Finally, in order to further test these scaling relations, it would also be interesting to study higher dimensions for which the scaling exponents are different.

\ack
The author specially thanks to L.F. Cugliandolo for stimulating discussions and suggestions. The author also thanks to E. Katzav and M. Pleimling for interesting discussions. Financial support from the Swiss National Science Foundation under MaNEP and Division II is also acknowledged.

\nosections


\begin{thebibliography}{99}

\bibitem{magnetic}
Lemerle S, Ferr\'e J, Chappert C, Mathet V, Giamarchi T and Le Doussal P, 1998 \PRL {\bf 80} 849
\nonum
Repain V, Bauer M, Jamet JP, Ferr\'e J, Mougin A, Chappert C and Bernas H, 2004 {\it Europhys. Lett.} {\bf 68} 460
\nonum
Bauer M, Mougin A, Jamet JP, Repain V, Ferr\'e J, Stamps RL, Bernas H and Chappert C, 2005 \PRL {\bf 94} 207211
\nonum
Metaxas PJ, Jamet JP, Mougin A, Cormier M, Ferr\'e J, Baltz V, Rodmacq B, Dieny B and Stamps RL, 2007 Creep and flow regimes of domain wall motion in ultrathin Pt/Co/Pt films with perpendicular anisotropy {\it Preprint} arXiv:cond-mat/0702654.

\bibitem{ferroelectric}
Tybell T, Paruch P, Giamarchi T and Triscone JM, 2002 \PRL {\bf 89} 097601
\nonum
Paruch P, Giamarchi T and Triscone JM, 2005 \PRL {\bf 94} 197601

\bibitem{rusos}
Blatter G, Feigel'man MV, Geshkenbein VB, Larkin AI and Vinokur VM, 1994 {\it Rev. Mod. Phys.} {\bf 66} 1125
\nonum
Nattermann T and Scheidl S, 2000 {\it Adv. Phys.} {\bf 49} 607

\bibitem{vortex}
Du X, Li G, Andrei EY, Greenblatt M and Shuk P, 2007 {\it Nature Phys.} {\bf  3}, 111

\bibitem{us}
Bustingorry S, Cugliandolo LF and Dom\'{\i}nguez D, 2006 \PRL {\bf 96} 027001
\nonum
Bustingorry S, Cugliandolo LF and Dom\'{\i}nguez D, 2007 \PR B {\bf 75} 024506

\bibitem{grains}
Foiles SM and Hoyt JJ, 2006 {\it Acta Mater.} {\bf 54} 3351
\nonum
Trautt ZT, Upmanyu M and Karma A, 2006 {\it Science} {\bf 314} 632

\bibitem{Barabasi-Stanley}
Barab\'asi A-L and Stanley HE, 1995 {\it Fractal
concepts in surface growth} (Cambridge: Cambridge University Press)
\nonum
Halpin-Healey T and Zhang Y-C, 1995 {\it Phys. Rep.} {\bf 254} 215

\bibitem{cracks}
Bouchaud E, 1997 \JPCM {\bf 9} 4319
\nonum
Ponson L, Bonamy D and Bouchaud E, 2006 \PRL {\bf 96} 035506
\nonum
Alava M, Nukalaz PKVV and Zapperi S, 2006 {\it Adv. Phys.} {\bf 55} 349

\bibitem{Cukupa}
Cugliandolo LF, Kurchan J and Parisi G, 1994 {\it J. Phys.} I {\bf 4} 1641

\bibitem{Pleimling}
R\"othlein A, Baumann F and Pleimling M, 2006 \PR E {\bf 74} 061604

\bibitem{cuentas_EW}
Bustingorry S, Iguain JL and Cugliandolo LF, {\it Out-of-equilibrium relaxation of the Edwards-Wilkinson elastic line}, 2007 {\it J. Stat. Mech} at press.

\bibitem{KPZ}
Kardar M, Parisi G and Zhang YC, 1986 \PRL {\bf 56} 889

\bibitem{Spaniards}
Ramasco JJ, Lopez JM and Rodriguez MA, 2006 {\it Europhys. Lett.} {\bf 76} 554

\bibitem{Bramwell}
Bramwell ST, Holdsworth PCW and Pinton JF, 1998 {\it Nature} {\bf 396} 552

\bibitem{Racz}
R\'acz Z, 2003 {\it SPIE Proceedings} {\bf 5112} 248

\bibitem{Chamon-etal}
Chamon C, Kennett MP, Castillo HE and Cugliandolo LF, 2002 \PRL {\bf 89} 217201
\nonum
Castillo HE, Chamon C, Cugliandolo LF and Kennett MP, 2002 \PRL {\bf 88} 237201
\nonum
Castillo HE, Chamon C, Cugliandolo LF, Iguain JL and Kennett MP, 2003 \PR B {\bf 68} 134442
\nonum
Chamon C, Charbonneau P, Cugliandolo LF, Reichman D and Sellitto M, 2004 {\it J. Chem. Phys.} {\bf 121} 10120

\bibitem{Chamon-Cugliandolo}
Chamon C and Cugliandolo LF, 2007 {\it J. Stat. Mech} P07022

\bibitem{Aarao}
Aar\~ao Reis FDA, 2004 \PR E {\bf 72} 032601
\nonum
Paiva T and Aar\~ao Reis FDA, 2007 {\it Surf. Sci.} {\bf 601} 419
\nonum
Oliveira TJ and Aar\~ao Reis FDA, 2007 Finite-size effects in roughness distribution scaling {\it Preprint} arXiv:0706.1307.

\bibitem{Marinari}
Marinari E, Pagnani A, Parisi G and R\'acz Z, 2002 \PR E {\bf 65} 026136

\bibitem{static-inter}
Foltin G, Oerding K, R\'acz Z, Workman RL and Zia RKP, 1994 \PR E {\bf 50} R639
\nonum
Plischke M, R\'acz Z and Zia RKP, 1994 \PR E {\bf 50} 3589
\nonum
R\'acz Z and Plischke M, 1994 \PR E {\bf 50} 3530

\bibitem{Zoltan}
Antal T and R\'acz Z, 1996 \PR E {\bf 54} 2256

\bibitem{Rosso}
Rosso A, Krauth W, Le Doussal P, Vannimenus J and Wiese KJ, 2003 \PR E {\bf 68} 036128

\bibitem{gyorgyi}
Antal T, Droz M, Gyorgyi G and Racz Z, 2001 \PRL {\bf 87} 240601
\nonum
Gyorgyi G, Moloney NR, Ozogany K and Racz Z, 2007 \PR E {\bf 75} 021123

\bibitem{Bustingorryetal}
Bustingorry S, Iguain JL, Chamon S, Cugliandolo LF and Dom\'{\i}nguez D, 2006 {\it Europhys. Lett.} {\bf 76} 856

\bibitem{Lam}
Lam C-H and Shin FG, 1998 \PR E {\bf 57} 6506
\nonum
Lam C-H and Shin FG, 1998 \PR E {\bf 58} 5592
\nonum
Buceta RC, 2005 \PR E {\bf 72} 017701

\bibitem{Leto}
Cugliandolo LF, 2004 {\it Slow Relaxations and Nonequilibrium Dynamics in Condensed Matter} ({\it Les Houches-Ecole d'Ete de Physique Theorique vol 77}), ed J-L Barrat {\it et al.} (Berlin: Springer) Also available as [cond-mat/0210312]

\bibitem{Favi}
Family F and Vicsek T, 1985 \JPA {\bf 18} L75

\bibitem{Kolton}
Kolton A, Rosso A and Giamarchi T, 2005 \PRL {\bf 95} 180604

\bibitem{Eytan}
Katzav E and Schwartz M, 2004 \PR E {\bf 69} 052603
\nonum
Pr\"ahofer M and Spohn H, 2004 {\it J. Stat. Phys.} {\bf 115} 255
\nonum
Colaiori F and Moore MA, 2001 \PR E {\bf 65} 017105
\nonum
Colaiori F and Moore MA, 2001 \PR E {\bf 63} 057103

\bibitem{Barrat}
Barrat J-L and Kob W, 1999 \JPCM {\bf 11} 247
\nonum
Barrat J-L and Kob W, 1999 {\it Europhys. Lett.} {\bf 46} 637

\bibitem{Bonn}
Bonn D, Tanaka H, Wegdam G, Kellay H and Meunier J, 1999 {\it Europhys. Lett.} {\bf 45} 52
\bibitem{Tanaka}
Tanaka H, Jabbari-Farouji S, Meunier J and Bonn D, 2005 \PR E {\bf 71} 021402

\end{thebibliography}
\end{document}